\documentclass[twocolumn,superscriptaddress,preprintnumbers,amsmath,amssymb,floatfix]{revtex4}
 
\usepackage{graphicx}
\usepackage{dcolumn}
\usepackage{bm}
\usepackage{subfigure}
\usepackage{amsmath}
\usepackage{wasysym} 
\usepackage{times}

\newcommand{\comment}[1]{}   

\newcommand{\newsim}{{\raise.17ex\hbox{$\scriptstyle\sim$}}}

\begin{document}

\title{{ \Large Magnetometry with Mesospheric Sodium}}

\author{J.\ M.\ Higbie}
\email{james.higbie@bucknell.edu}
\affiliation{Department of Physics and Astronomy, Bucknell University,
Lewisburg, PA 17837
}
\author{S.\ M.\ Rochester}\affiliation{Department of Physics, 
University of California, Berkeley, CA  94720-7300}
\author{B.\ Patton}\affiliation{Department of Physics, 
University of California, Berkeley, CA  94720-7300}
\author{R.\ Holzl\"{o}hner} \affiliation{European Southern Observatory (ESO), Garching bei M\"{u}nchen,
D-85748, Germany }
\author{D.\ Bonaccini Calia}  \affiliation{European Southern Observatory (ESO), Garching bei M\"{u}nchen,
D-85748, Germany }
\author{D.\ Budker}
\email{budker@berkeley.edu} \affiliation{Department of
Physics, University of California, Berkeley, CA 94720-7300}
\affiliation{Nuclear Science Division, Lawrence Berkeley National
Laboratory, Berkeley CA 94720}

\date{Dec. 18, 2009}

\begin{abstract}
Measurement of magnetic fields on the few-hundred-kilometer 
length scale is significant for a variety of geophysical applications including mapping of 
crustal magnetism and ocean-circulation measurements, 
yet  available techniques for such measurements are very expensive or of limited accuracy.
We propose a scheme for remote detection of  magnetic fields 
using the naturally occurring atomic-sodium-rich layer in the mesosphere and existing 
high-power lasers developed for laser guide-star applications. 
The proposed scheme offers dramatic reduction in cost, opening the way to 
large-scale magnetic
mapping missions.
\end{abstract}
\maketitle

\section{Introduction}
Measurements of geomagnetic fields are an important tool for peering
into the earth's interior, 
with measurements at differing spatial scales giving information about
sources at corresponding depths.  Measurements of fields on the
few-meter scale can locate buried ferromagnetic objects (e.g.
unexploded ordnance or abandoned vessels containing toxic waste),
while maps of magnetic fields on the kilometer scale are used to
locate geological formations promising for mineral or oil extraction.
On the largest scale, the earth's dipole field gives information about
the geodynamo at depths of several thousand kilometers.  Magnetic
measurements at intermediate length scales, in the range of several
tens to several hundreds of kilometers likewise offer a window into
important scientific phenomena, including the behavior of the outer
mantle, the solar-quiet dynamo in the ionosphere \cite{Campbell1989},
and ionic currents as probes of ocean
circulation\cite{TylerOcean2003}, a major actor in models of climate
change.  

To avoid contamination from local perturbations,
magnetic-field measurements on this length scale must typically be
made at a significant height above the earth's surface.  Though
magnetic mapping at high altitude has been realized with
satellite-borne magnetic sensors
\cite{Friis2006,Slavin2008,Purucker2007}, the great expense of
multi-satellite missions places significant limitations on their
deployment and use.  Here we introduce a high-sensitivity method of
measuring magnetic fields with 100-km spatial resolution without the
cost of spaceborne apparatus, by exploiting the naturally occurring
atomic sodium layer in the mesosphere and the significant
technological infrastructure developed for astronomical laser guide
stars (LGS). 
This method promises to enable creation of global sensor arrays for
continuous mapping and monitoring of geomagnetic fields without
interference from ground-based sources.

\begin{figure}
\includegraphics[width=7.5cm]{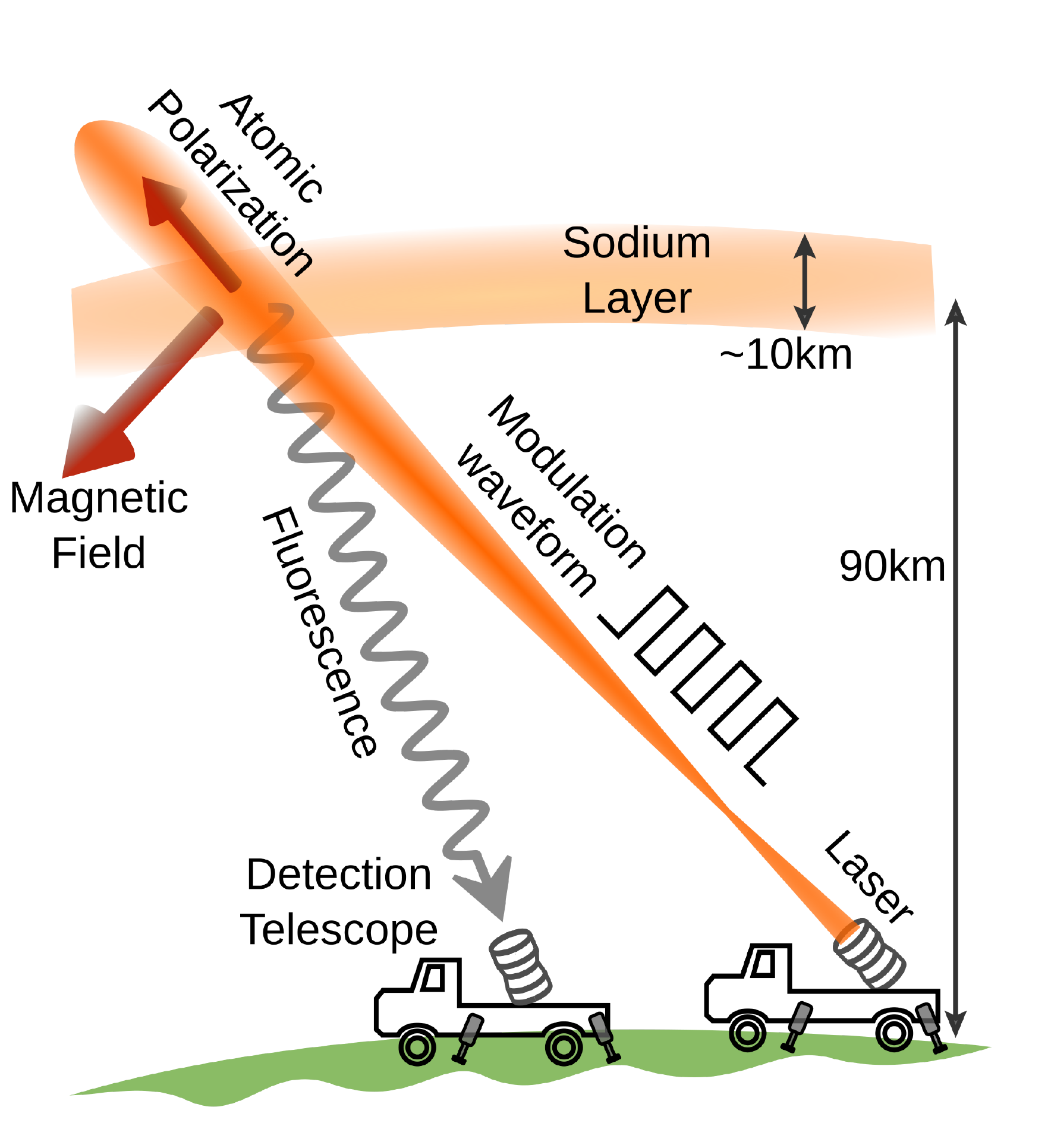}  
\caption{\label{fig:scheme}
  Fluorescent detection of magneto-optical resonance of mesospheric
  sodium.
 (Diagram not to scale). Circularly polarized laser light at
  $589\,$nm, modulated near the Larmor frequency, pumps atoms in the
  mesosphere. The resulting spin polarization (pictured as
  instantaneously oriented along the laser beam propagation direction)
  precesses around the local magnetic field.  Fluorescence collected
  by a detection telescope exhibits a resonant dependence on the
  modulation frequency.}
\end{figure} 

\section{Principle of the Method}
The measurement we envisage is closely related to the techniques of
atomic magnetometry, appropriately adapted to the conditions of the
mesosphere.  The principle of the method is to measure the magnetic
field using spin precession of sodium atoms by creating atomic spin
polarization, allowing it to evolve coherently in the magnetic field,
and determining the post-evolution spin state. Conceptually, these
processes are distinct and sequential, although in practice they may
occur simultaneously and at different times for different atoms.
Preparation of spin-polarized mesospheric sodium atoms is achieved by
optical pumping, as was proposed for mesospheric sodium in the seminal
paper on sodium LGS by Happer et al.~\cite{Happer1994}.  In the
simplest realization, the pumping laser beam is circularly polarized
and is launched from a telescope at an angle approximately
perpendicular to the local magnetic field, as shown in Fig.\
\ref{fig:scheme}.  In the presence of the magnetic field, the
transverse polarization generated by optical pumping precesses around
the magnetic-field direction at the Larmor frequency.  In order to
avoid ``smearing'' of the atomic polarization due to this precession,
the optical-pumping rate is modulated near the Larmor frequency, as
first demonstrated by Bell and Bloom \cite{BellBloom1961}. When the
modulation frequency and the Larmor frequency coincide, a resonance
results, and a substantial degree of atomic polarization is obtained.
The atomic polarization in turn modifies the fluorescence from the
sodium atoms, which can be detected by a ground-based telescope.  This
allows the sodium atoms to serve as a remote sensor of the magnitude
of the magnetic field in the mesosphere, i.e. as a scalar
magnetometer.  By stationing lasers and detectors on a
few-hundred-kilometer grid, a simultaneous map of magnetic fields may
be obtained; alternatively, the laser and detector can be mounted on a
relocatable stable platform such as a ship or truck to facilitate magnetic
surveying.  In contrast to both ground-based and satellite-based
measurements, the platform is not required to be magnetically clean or
quiet.

The magneto-optical resonance manifests itself as a sharp increase in
the returned fluorescence for the D2 line of sodium, or a decrease for
the D1 line, as a function of the modulation frequency.  The sodium
fluorescence may be captured either by a single-channel photodetector
or by an imaging array (e.g., a CCD camera); the latter allows
additional background discrimination and the possibility of separately
analyzing fluorescence signals from different altitudes if the imaging
and laser-beam axes are not parallel.

\section{Measurement Parameters}

The magnetometric sensitivity of this technique is governed by the
number of atoms involved in the measurement, the coherence time of the
atomic spins, and the fraction of the total fluorescence intercepted
by the detector, each of which we seek to maximize.  A detailed
discussion of mesospheric properties relevant to LGS has recently been
given \cite{Holzloehner2009}.  Briefly, the altitude of the sodium
layer is $\newsim90\,$km, its thickness is $\newsim10\,$km, and sodium
is present at a typical temperature of $180\,$K and a number density
of around $3\times 10^9\,{\rm m}^{-3}$.  This density is low by
vapor-cell standards, but the interaction volume and hence the atom
number can be quite large, limited chiefly by available laser power.
The coherence time is limited primarily by collisions with other
atmospheric molecules and secondarily by atom loss from the region
being probed (e.g., due to diffusion or wind).  A velocity-changing
collision occurring after an atom is pumped typically removes the atom
from the subset of velocity classes which are near-resonant with the
laser light; as a consequence, these collisions result in an effective
decay of spin polarization.  Moreover, spin-exchange collisions of
sodium atoms with unpolarized paramagnetic species in the mesosphere,
predominantly ${\rm O}_2$, result in a randomization of the electron
spin, and therefore also lead to decay of sodium polarization.  To our
knowledge, the spin-exchange cross-section of oxygen with sodium has
not been measured, and in fact it is anticipated that measurement of
mesospheric coherence times will more tightly constrain this important
LGS parameter.  However, its magnitude can be estimated from other
known spin-exchange cross-sections, leading to an expected
spin-damping time on the order of $500\,\mu$s.  The fraction of
intercepted fluorescence is determined by the solid angle subtended by
the detection telescope and the angular emission pattern of the
fluorescing atoms; for a $1$-m$^2$ telescope and isotropic emission,
the fraction is approximately $10^{-11}$ when the detector is directly
below the fluorescing sample.

Sodium atoms in the mesosphere are distributed both in position and in
velocity. They can be conceptually divided into velocity classes and
spatial volumes, in each of which the collection of atoms functions
approximately 
as an independent magnetometer. The width in velocity $\Delta v$ of an
individually addressable velocity class is set by the Doppler effect
and the natural linewidth $\gamma_0\approx 2\pi \times 10\,$MHz of the
sodium transition.  
We can thus approximately calculate 
the sensitivity by calculating the evolution of a
single sub-volume and velocity class; we then
determine the total fluorescence by summing over sub-volumes and
velocity classes.
  This approximate treatment is valid for length scales larger
than the typical distance traveled by atoms in one coherence time, and
for velocity classes that are approximately uncoupled. The latter
condition requires that the laser intensity be low, so that radiation
pressure is small, and that there be negligible probability of an atom
begin transfered from one near-resonant velocity class to another by
velocity-changing collisions.

The atom number can be maximized either by increasing the interaction
volume or by including more velocity classes.  Since the most probable
atomic velocity component along the laser beam propagation direction
is zero, broadening the laser spectrum to include velocity classes
away from zero velocity will offer diminishing returns as spectral
widths approaching the Doppler linewidth ($\newsim1\,$GHz) are reached.
By contrast, defocusing the laser beam to illuminate a larger region
in space suffers no limitation other than the amount of laser power
that can be supplied at a reasonable cost. For this reason, as well as
for simplicity of implementation, defocusing the laser beam appears
advantageous. 
If the observed relaxation due to velocity-changing collisions is
significantly more rapid than that due to spin-exchange collisions,
however, it may prove beneficial to broaden the laser spectrum to a
width comparable to the Doppler linewidth, so that velocity-changing
collisions will no longer cause loss of polarization but will merely
cause atoms to be pumped and probed in different velocity classes.

\section{Results} 
\begin{figure} 
\includegraphics[width=8.5cm]{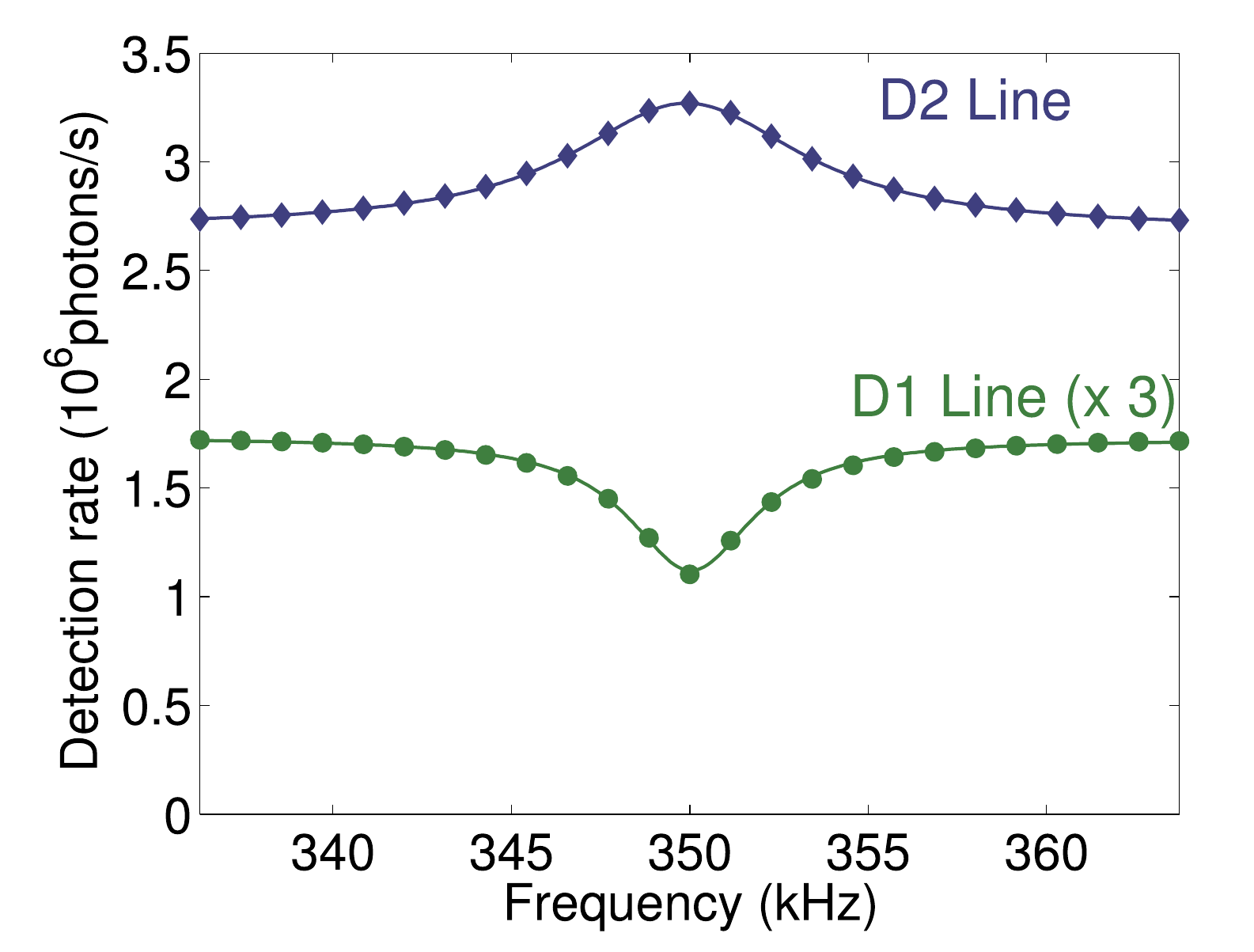} 
\caption{\label{fig:resonancesHF}
Calculated magneto-optical resonance profiles for mesospheric sodium.
The resonances shown correspond to the D2 $F$=2 $\rightarrow$ $F'$=3
 (upper curve, blue diamonds) and D1 $F$=2 $\rightarrow$ $F'$=1 (lower 
curve, green circles) sodium lines. The D1 curve has been multiplied by a factor of three to improve 
visibility.
Symbols are the results of numerical calculations, and solid lines are Lorentzian fits to these results.
Calculations are for an intensity $I=1\,{\rm W}/{\rm m}^2$ and a modulation duty cycle of $20$\% with
  a detector collection area of $1\,{\rm m}^2$. 
}
\end{figure}

\begin{figure}  
\includegraphics[width=8.5cm]{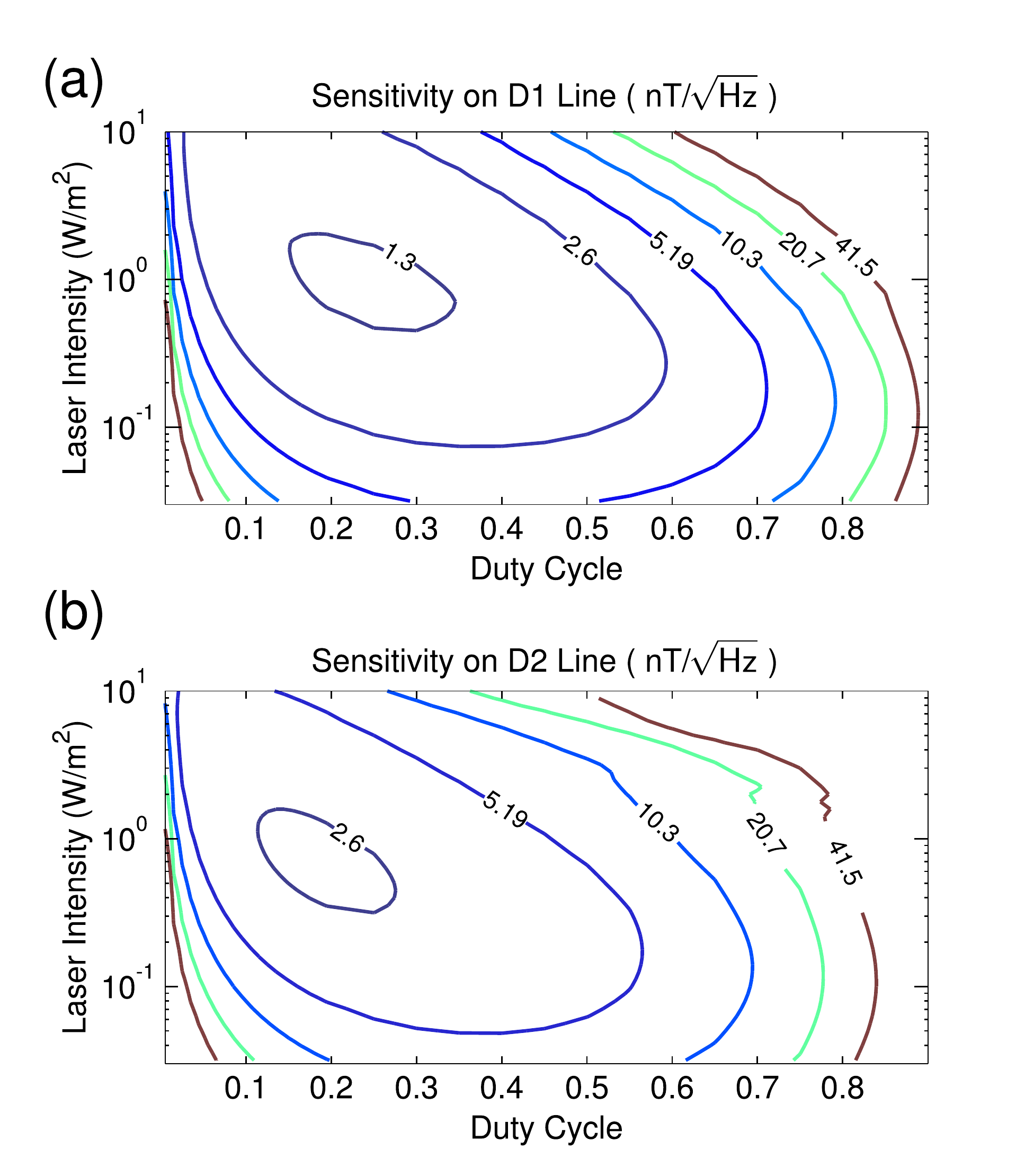}
\caption{\label{fig:contoursHF} 
Contour plot of  
calculated magnetometer sensitivity
 as a function of pumping duty cycle and intensity.
 Sensitivity is 
calculated including the hyperfine effect for the D2 $F$=2 $\rightarrow$ $F'$=3 and D1
$F$=2$\rightarrow F'$=1 lines, using a detector area of $1\,{\rm m}^2$,
a spin-exchange collision time of $500\,\mu$s,
and a velocity-changing collision time of $200\,\mu$s.
Contours are logarithmically spaced at intervals of one octave.}
\end{figure}

One expects on intuitive grounds that the optimum laser intensity
resonant with a single velocity class should be such that the
characteristic rate of optical pumping $\Gamma_p \equiv {\gamma_0}
{I}/2{I_{\rm sat}}$ (where $I$ is the laser intensity and $I_{\rm
  sat}\approx 60\,{\rm W}/{\rm m}^2$ is the saturation intensity of
the sodium cycling transition) is on the same order as the decay rate
of atomic polarization.  For higher intensities, optical pumping will
``reset'' the precession before its coherence time has been fully
exploited, while use of lower intensities sacrifices signal without
improving the coherence time.  We have performed a detailed
ground-state density-matrix analysis of spin precession and optical
pumping on the D1 and D2 transitions of sodium. In this analysis, we
employ a circularly polarized pump laser beam oriented at right angles
to a magnetic field of $0.5\,$G, with a spin-exchange collision time
of $500\,\mu$s and a velocity-changing collision time of $200\,\mu$s.
Since we employ optical intensities substantially lower than the
saturation intensity, we expect the ground-state method to be
accurate; we have also performed calculations
of the resonance contrast for selected
parameters 
using the full (ground and    
excited-state) optical Bloch equations. 
For a variety of settings of the pump light intensity and of
the duty cycle (defined as the duration of a single pumping light
pulse divided by the period of the modulation), we calculate resonance
spectra as functions of the laser modulation frequency. Sample spectra
for the laser tuned to the D1 $F=2 \rightarrow F'=1$ and the D2 $F=2
\rightarrow F'=3$ lines are shown in Fig.\ \ref{fig:resonancesHF}.
From the width and peak height of these spectra, as well as the optical 
shot noise of the detected fluorescence, we calculate the
magnetometric sensitivity. We assume that the noise is dominated by the fundamental
shot noise and therefore neglect technical noise due to the photometric measurement.
Contour plots of the sensitivity are shown
as functions of the duty cycle and the laser intensity in Fig.\
\ref{fig:contoursHF} for the D1 and D2 lines.
In these calculations, a fixed number of velocity classes is
considered (specifically fifteen, or five per excited-state hyperfine
component), and the effective laser beam size is adjusted to maintain
a constant launched laser power of $20\,$W (typical of the latest
generation of LGS lasers), which results in different numbers of
participating atoms for different laser intensities.  The optimum
sensitivity of $1.2\, {\rm nT/\sqrt{Hz}}$ occurs on the D1 transition,
at a duty cycle of $25\%$ and pump intensity during each pulse of
$1.0\, {\rm W/ m}^2$. The optimum intensity corresponds to an
effective laser beam diameter in the mesosphere of around $5\,$m. This
sensitivity is approximately one order of magnitude worse than the
limit set by quantum spin-projection noise for the sodium atoms,
presumably as a result of hyperfine structure and Doppler broadening.
The optimum on the D1 line offers superior magnetometric sensitivity
in part because the D1 resonances are dark, i.e., they result in a
reduction of fluorescence, so that the photon shot noise is smaller
and broadening of the resonance is reduced. If technical rather than
fundamental noise sources dominate, then the D2 resonance may be
preferable for its larger signal size.

We note also that the time scale of velocity-changing collisions is
comparable to the relaxation time of the atomic spins, so that it is
likely for a given atom to interact more than once with the laser
light while remaining in a single velocity class.  Thus the
excited-state hyperfine structure is resolved throughout the pumping
and probing process, and proximity to a given ground-to-excited
hyperfine transition can strongly influence the light-atom
interaction\cite{Auzinsh2009}.  For the same reason, magneto-optical
resonances involving higher polarization multipoles such as alignment
(which can be prepared by pumping with linearly polarized light) are
expected to be observable in mesospheric sodium.

\section{Additional Effects}
The magneto-optical resonance linewidth in mesospheric sodium is
broader than in typical vapor-cell magnetometers; as a result, several
effects important in vapor cells are less significant for
mesospheric-sodium measurements.  The quadratic Zeeman shift, for
instance, leads to a splitting of the resonance into multiple
resonances spaced by $\,\newsim150\,$Hz. However, since the width of
the resonance at the optimum is around $5\,$kHz, this splitting will
merely result in a small (and calculable) distortion of the lineshape.
The natural inhomogeneity of the geomagnetic field will also affect
the measurement, since the resonant frequency varies with altitude.
Variation of the earth's field over the $10$-km height of the sodium
layer is on the order\cite{NOAAMagCalc} of $100\,$nT, corresponding to
about $700\,$Hz in Larmor frequency.  Thus, although the modulation
pumping laser cannot perfectly match the resonance frequency
throughout the height of the mesosphere, this inhomogeneous broadening
is again small compared to the width of the resonance at the
sensitivity optimum.  Consequently, we expect the effects of both the
natural magnetic gradient and the quadratic Zeeman shift to be small.
Temporal variations of the magnetic field in the mesosphere are, in
principle, merely part of the signal being measured, and not an
instrumental limitation.  However, large enough fluctuations could
make it difficult to track the resonance frequency. We take as a
likely upper bound for the magnetic fluctuations on time-scales of
$1\,$s to $100\,$s the typical observed value at the earth's surface
under ordinary conditions of around $1\,$nT. As this is again
substantially smaller than the resonance linewidth, we expect that
except during magnetic storms, it should not be difficult to keep the
laser modulation frequency on resonance. Variations of the height and
density of the sodium layer itself are an additional practical
concern. A realistic measurement will require reducing sensitivity to
such variations through comparison of on-resonant and off-resonant
signals, either by temporally dithering the modulation frequency or by
employing spatially separated pump beams with different modulation
frequencies.

A further deviation of the real experiment from the idealization
embodied in the calculations comes from turbulence in the lower
atmosphere, which causes  random phase shifts 
in distinct transverse patches of the
laser beam.  In the typical LGS
application, the far-field diffraction (or speckle) pattern in the mesosphere from
these low-altitude phase patches consists of elongated filaments whose individual lateral
size is set by the numerical aperture of the laser launch telescope,
but whose collective extent is governed by the size of the atmospheric
patches\cite{Happer1994}.  Fluctuation of these filaments in time results in
undesirable effects including random changes in pump-laser intensity
in the mesosphere, motion of the illuminated column, and variation of
the returned fluorescence.  The relatively low intensity and
large beam area indicated
by our magnetometry calculations make such filamentation a lesser
concern, however, since the
beam diameter may be kept
within the Fried length\cite{Fried1966} of around $0.1\,$m
in the most turbulent region of the lower atmosphere.   
Although lensing and beam-steering
due to atmospheric variations will prevent precise fine-tuning of
laser intensity, we do not anticipate that such variations will
strongly affect the sensitivity obtainable with the proposed
technique. We plan nevertheless to perform detailed modeling of
atmospheric effects using physical optics, as has recently been done
with reference to LGS \cite{holzlohner2008}.

\section{Conclusions}
In conclusion, we have presented a promising alternative to satellite missions for 
the measurement of geomagnetic fields. 
The proposed method requires only ground-based apparatus, and is consequently 
substantially less expensive per sensor
than  a satellite formation, while still achieving high magnetometric sensitivity. 
 We anticipate that the low cost of deployment will make possible 
large-scale magnetic mapping and monitoring applications
 at the $100$-km length scale, with temporal and spatial 
coverage that would be difficult to obtain by current techniques.
Furthermore, as satellites cannot be operated as low as $100\,$km in altitude without 
excessive drag and heating, remotely-detected mesospheric magnetometry
promises
superior spatial resolution of terrestrial sources.   
In addition, the technique offers to supplement
 existing ground-based magnetic
observatory data, allowing high-precision magnetic monitoring from a mobile platform without
the requirement of a large-area, remote, and magnetically clean observation site on the earth's surface.
We are currently  constructing a $20$-W-class laser projection system and working  to 
implement this technique in proof-of-principle magnetic-field
measurements.
 
{
The authors acknowledge stimulating discussions 
with Peter Milonni, William Happer, Michael Purucker, and Stuart Bale. 
This work is supported by 
the NGA NURI program.
}

 \bibliographystyle{prsty}


\bibliography{general} 

\end{document}